\documentclass[twoside]{dis07}
\usepackage[latin1]{inputenc}
\usepackage[dvips]{graphicx,epsfig,color}
\usepackage{wrapfig,rotating}
\usepackage{amssymb,amsmath,array}

\begin{document}
\title{Early Electroweak and Top Quark Physics with CMS}

%***********************************************************************
% AUTHORS INFORMATION AREA
%***********************************************************************
\author{Frank-Peter Schilling 
\thanks{
%Presented at DIS 2007
on behalf of the CMS Collaboration}
\vspace{.3cm}\\
University of Karlsruhe - Institut f\"ur Experimentelle Kernphysik \\
D-76128 Karlsruhe - Germany
}
%***********************************************************************
% END OF AUTHORS INFORMATION AREA
%***********************************************************************

\maketitle

\begin{abstract}

The Large Hadron Collider is an ideal place for precision measurements
of the properties of the electroweak gauge bosons $W^\pm,Z^0$, as
well as of the top quark. In this article, a few highlights of the
prospects for performing such measurements with the CMS detector are
summarized, with an emphasis on the first few 1/fb of data.

\end{abstract}

\section{Introduction}

At the Large Hadron Collider (LHC), $W^\pm$ and $Z^0$ bosons as well
as top quarks will be produced copiously, due to the large
center-of-mass energy of 14 TeV (which leads to increased production
cross sections with respect to e.g. the TEVATRON) as well as the high
luminosity of up to $10^{34} \ {\rm cm}^{-2} {\rm s}^{-1}$.  These
samples can be used not only for precision measurements of standard
model parameters such as $m_W$ and $m_t$, but also for detector
commissioning, alignment and calibration. Furthermore, standard model
processes involving $W^\pm,Z^0$ bosons and top quarks constitute the
primary sources of background in many Higgs boson and new physics
searches.

This article~\cite{url} summarizes a few highlights of recent
studies~\cite{ptdr2} of the potential of the CMS experiment regarding
top quark and electroweak physics, in particular in view of the first
few 1/fb of data.  They have been performed with a full detector
simulation, are based on the reconstruction software and calibration
procedures demonstrated in~\cite{ptdr1}, and include estimates of the
main systematic uncertainties.

%%%%%%%%%%%%%%%%%%%%%%%%%%%%%%%%%%%%%%%%%%%%%%%%%%%%%%%%%%%%%%%%%%%%%%%%%%%%%%%

\section{Electroweak Physics}

The reactions $pp\rightarrow W+X$ and $pp\rightarrow Z+X$, with
subsequent leptonic decays of the $W^\pm$ and $Z^0$ bosons, have a
large cross section and are theoretically well understood. Cross
sections above 10 nb (1 nb) are expected at the LHC for the
$W\rightarrow l+\nu$ ($Z\rightarrow l^++l^-$) channel in the fiducial
region of the CMS detector.  Thousands of leptonic $W^\pm$ and $Z^0$
decays will be recorded for luminosities as low as $1 \rm\ pb^{-1}$.
Hence, they are useful for many purposes, including a precise
luminosity monitor, a high-statistics detector calibration and
alignment tool and to demonstrate the performance of the CMS
experiment. These reactions will be among the first to be measured at
the LHC.

The measurement of the inclusive production of $W^\pm$ and $Z^0$
bosons with CMS has been studied in~\cite{wzmu} and~\cite{wzel} for
the muon and electron decay channel, respectively.  The emphasis has
been put on a start-up oriented event selection with high
purity. Already for an integrated luminosity of $1 \rm\ fb^{-1}$, the
uncertainty in the measured cross section will be dominated by
systematics. In case of the muon channel, 
\begin{eqnarray*}
\Delta\sigma / \sigma
(pp\rightarrow Z+X\rightarrow \mu\mu +X) = 0.13 {\rm\ (stat.)} \pm
2.3 {\rm\ (syst.)} \ \%  \\
\Delta\sigma / \sigma (pp\rightarrow
W+X\rightarrow \mu\nu +X) = 0.04 {\rm\ (stat.)} \pm\ 3.3 {\rm\
(syst.)} \ \% \ ,
\end{eqnarray*}
where the systematic error is dominated by a $2\%$ uncertainty
originating from the modeling of the boson $p_T$ dependence, which
enters in the acceptance determination.  Another important source of
theoretical uncertainty is the imperfect knowledge of the parton
density functions (PDFs), which affects the absolute normalization by
$5-7\%$~\cite{wzmu}.
%comparable to the uncertainty in the
%next-to-leading order (NLO) cross section calculation. 
Unless more
precise PDF sets become available, this will be a limiting factor in
comparisons between experiment and theory and in luminosity
measurements via $W,Z$ counting. But the argument can also be turned
around: These processes can yield important PDF constraints, even
without very precise knowledge of the luminosity, in particular by
measuring the shapes of differential lepton distributions~\cite{mandy}.

\begin{figure}
\begin{minipage}{0.49\linewidth}
\centering
\epsfig{file=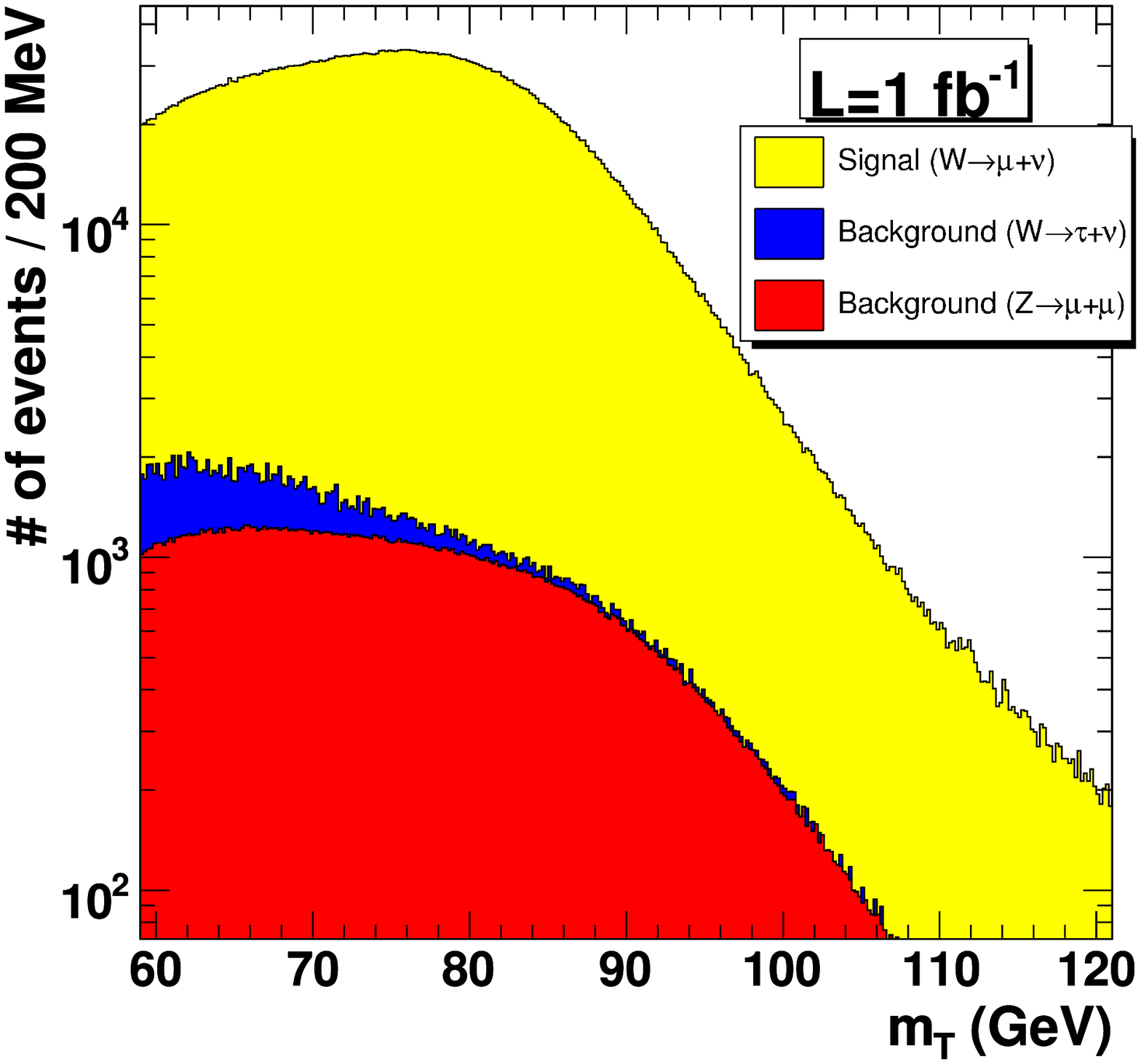,width=0.8\linewidth} %\\ (a)
\end{minipage}
\begin{minipage}{0.49\linewidth}
\centering
\epsfig{file=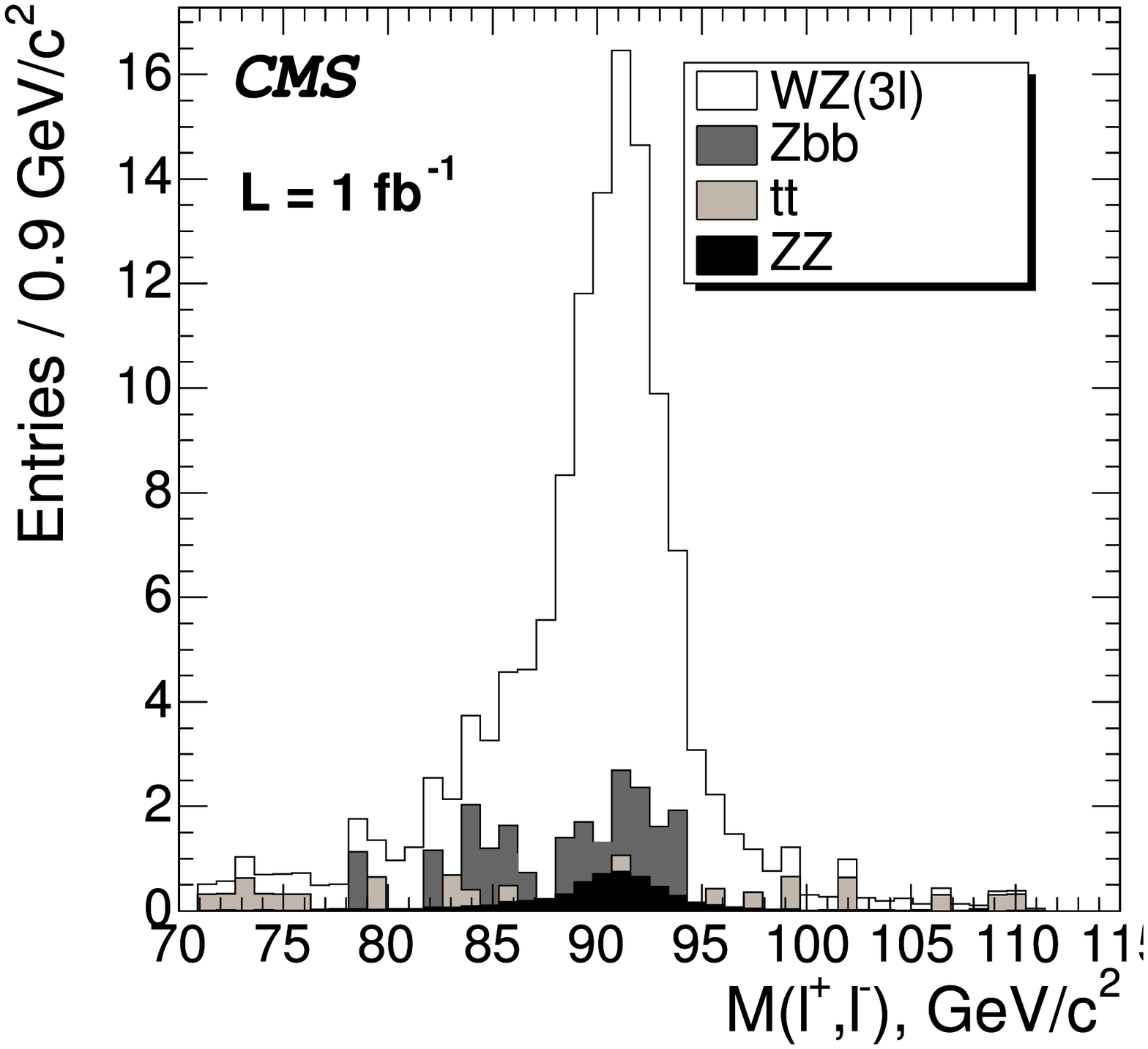,width=0.8\linewidth} %\\ (b)
\end{minipage}
\caption{(left) Transverse mass distribution in the $W\rightarrow \mu\nu$
channel for $1\rm\ fb^{-1}$. (right) Dilepton invariant mass distribution
in the $WZ\rightarrow 3l$ channel for $1\rm\ fb^{-1}$.}
\label{fig:wz}
\end{figure}

The $W^\pm$ boson mass is an important Standard Model (SM)
parameter. CMS has investigated the use of methods involving $W/Z$
ratios in the mass measurement, which have the advantage that many
experimental and theoretical uncertainties
cancel~\cite{gielekeller}. Figure~\ref{fig:wz}(left) shows the
simulated transverse mass distribution for $1 \rm\ fb^{-1}$ in the
muon channel~\cite{wmass}.  For both electron and muon channel, the
statistical error on $m_W$ is estimated as 40 (15) MeV for $1 \ (10)
\rm\ fb^{-1}$. The total experimental uncertainty is estimated as 40
(20) and 64 (30) MeV for the electron and muon channel, respectively.
Apart from the PDF uncertainty, the dominating theoretical uncertainty
originates from the modeling of the lepton $p_T$ distribution
(estimated as 30 MeV), which may be improved with higher-order
calculations.  Combining electron and muon channel, the uncertainty on
$m_W$ may be reduced to $10 {\rm\ (stat.)} \pm 20 {\rm\ (syst.)}$
for $10
\rm\ fb^{-1}$.

The production of diboson pairs can be used to probe triple gauge
boson couplings and thus the non-abelian gauge symmetry of electroweak
interactions. Such processes are also sensitive to new physics.  At
the LHC the production cross sections for $WZ$ and $ZZ$ pairs are
large (50 and 20 pb respectively). CMS has studied the production of
$WZ$ ($e$ or $\mu$ channels) as well as of $ZZ$ ($4e$ channel)
pairs~\cite{diboson}.  For $1 \rm\ fb^{-1}$, 97 events are expected in
the $WZ$ channel (Fig.~\ref{fig:wz}(right)), and a $5\sigma$ discovery is
possible with just $150 \rm\ pb^{-1}$ of data. In the $ZZ\rightarrow
4e$ channel, 71 events are expected for $10 \rm\ fb^{-1}$.  The large
signal over background (S/B) ratio makes these measurements very
useful to assess the background in the search for the Higgs boson.

%%%%%%%%%%%%%%%%%%%%%%%%%%%%%%%%%%%%%%%%%%%%%%%%%%%%%%%%%%%%%%%%%%%%%%%%%%%%%%%

\section{Top Quark Physics}

The $t\bar{t}$ production cross section at the LHC is $\sim 830 \rm\
pb$ (e.g.~\cite{topxsnlo}), which is more than two orders of magnitude
higher than at the TEVATRON. At a luminosity of $2*10^{33} \rm\
cm^{-2} s^{-1}$, about 1 $t\bar{t}$ pair will be produced per second,
predominantly gluon-induced. Also the cross section of the electroweak
production of single top quarks is large, $\sim 245 \rm\ pb$ in the
t-channel. In $1 \rm\ fb^{-1}$ of data, around 800K $t\bar{t}$ pairs
and 350K single top quarks will be produced, which makes the LHC
experiments ideal laboratories to precisely measure top quark
properties. In addition, since large samples of $t\bar{t}$ events will
be available already with the first year's data, they can also be used
as a detector commissioning tool, e.g. to study lepton identification
and isolation, jet and missing $E_T$ energy scales and b-tagging
performance.  The initial goal will be to measure the $t\bar{t}$ cross
section, followed by the mass measurement and studies of single top
production, polarization or search for flavor-changing neutral
currents (FCNC).

\begin{figure}
\begin{minipage}{0.49\linewidth}
\centering
\epsfig{file=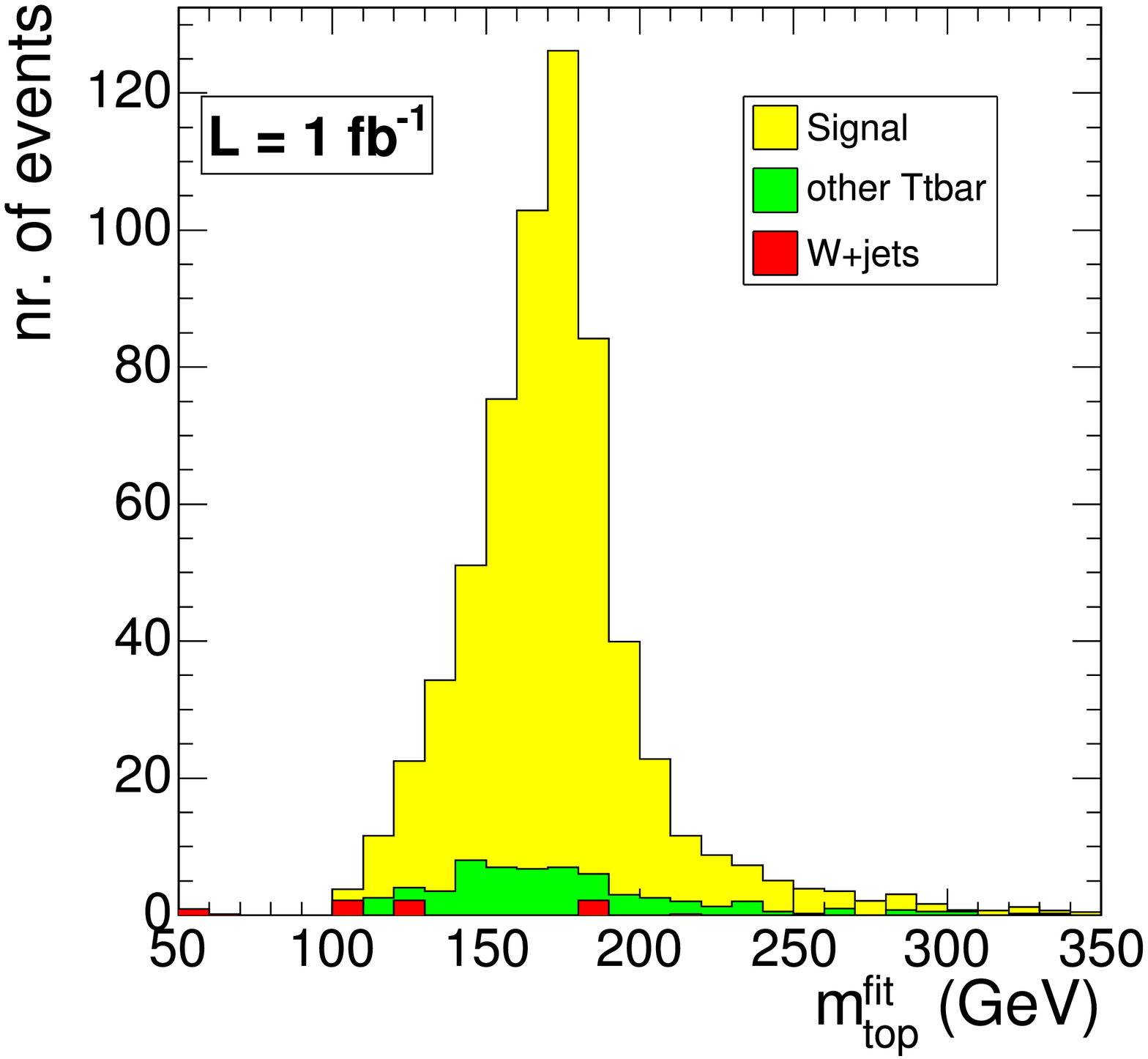,width=0.8\linewidth} %\\ (a)
\end{minipage}
\begin{minipage}{0.49\linewidth}
\centering
\epsfig{file=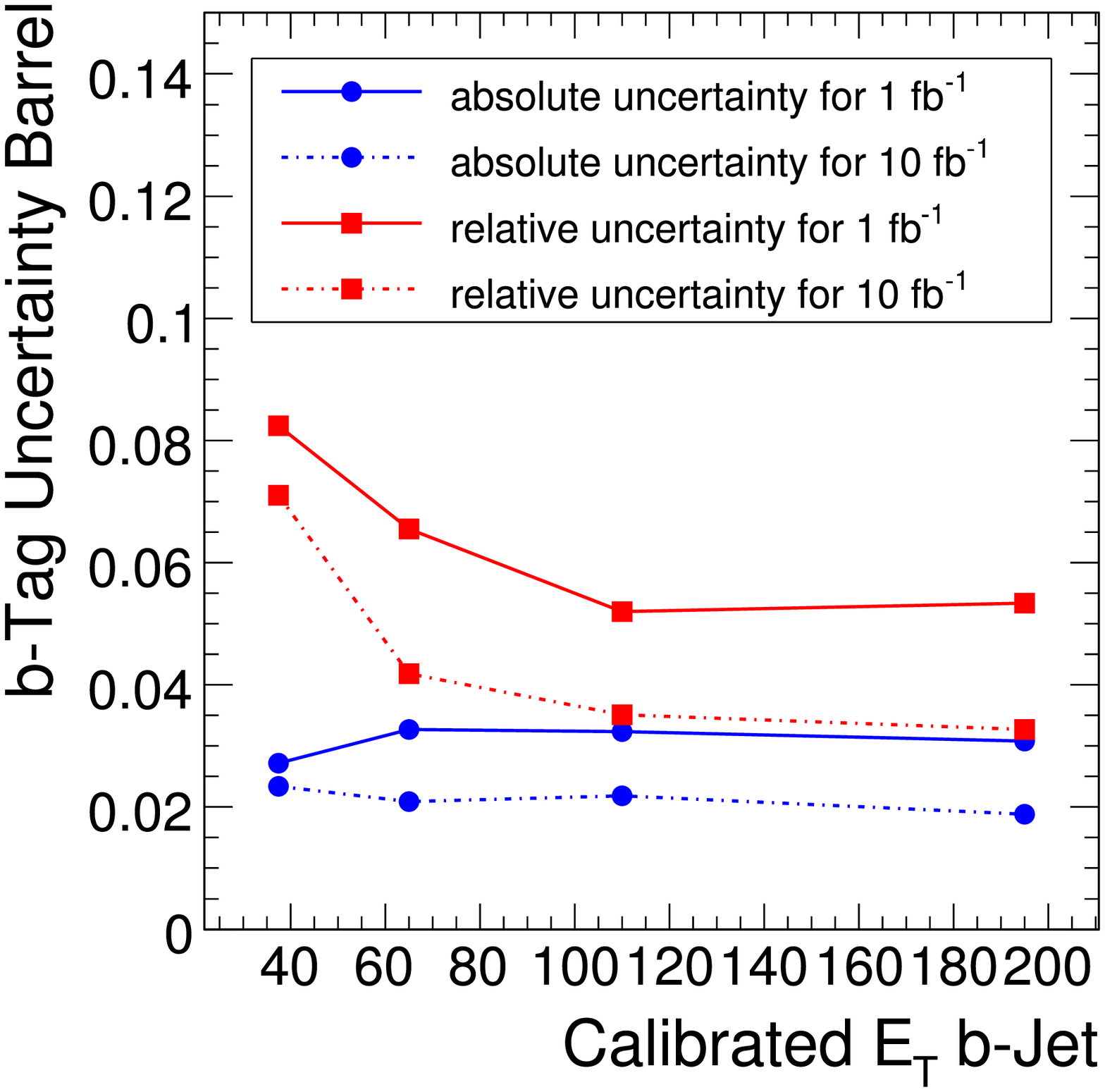,width=0.8\linewidth} %\\ (b)
\end{minipage}
\caption{(left) Reconstructed $m_t$ distribution
in the semileptonic channel for $1\rm\ fb^{-1}$. 
 (right) b-tagging uncertainty in the barrel detector as a function
of jet $p_T$, determined from $t\bar{t}$ events.
}
\label{fig:top}
\end{figure}

The measurement of the $t\bar{t}$ cross section has been studied in
all three decay modes~\cite{topxssemilep,topxsmassllhh}.  In the
semileptonic channel (Fig.~\ref{fig:top}(left)), the cross section can
be determined from event counting due to the high $S/B\sim 27$.  For
$1 \ (10) \rm\ fb^{-1}$, the statistical and systematic uncertainties
are estimated as $1.2 \ (0.4)$ and $9.2\%$ respectively, where the
systematic uncertainty is dominated by the knowledge of the b-tagging
efficiency, which is conservatively estimated as $7\%$. If it could be
reduced to $2\%$, the total error on $\sigma(t\bar{t})$ could be
reduced to $7\%$ at $10 \rm\ fb^{-1}$, which would already constrain
$m_t$ indirectly to $\Delta m_t \sim 2-3 \rm\ GeV$, comparable to the
precision of the direct measurements at the TEVATRON.  For the dilepton and
fully hadronic channels, the statistical (systematic) uncertainties
are estimated as $0.9 \ (11) \%$ and $3 \ (20) \%$ respectively at
$10 \rm\ fb^{-1}$.

The top quark mass  $m_t$ is related to the Higgs mass via loop corrections.
Also the measurement of $m_t$ has been studied in all decay modes.  In the
semileptonic channel~\cite{topmasssemilep}, a simple gaussian fit is
compared with the more sophisticated {\em ideogram} method. For $10
\rm\ fb^{-1}$, a precision of $\Delta m_t = 0.21 { \rm\ (stat.)} 
\pm 1.13 { \rm\ (syst.)}  \rm\ GeV$ is estimated for this
method. Thus, a 1 GeV uncertainty on $m_t$ looks achievable, but
requires a very good detector understanding.
The other decay modes~\cite{topxsmassllhh} have been investigated as
well.  In the dilepton channel an uncertainty of $\Delta m_t = 1.5 \
(0.5) { \rm\ (stat.)} \pm 2.9 \ (1.1) { \rm\ (syst.)} \rm\ GeV$ is
estimated for $1 (10) \rm\ fb^{-1}$, where the systematic error is
dominated by the jet energy scale uncertainty.  In the fully hadronic
channel, where a jet pairing likelihood is applied to improve the S/B
from 1/9 to 1/3 at constant efficiency, the estimate is $\Delta m_t =
0.6 { \rm\ (stat.)} \pm 4.2 { \rm\ (syst.)} \rm\ GeV$ for $1 \rm\
fb^{-1}$.

Due to the large cross section $t\bar{t}$ events are useful as a tool
to commission and calibrate the detector. For instance, a study has
shown that the light quark jet energy scale can be constrained to the
level of $3\%$ by applying a $m_W$ constraint in $t\bar{t}$
events~\cite{topjetscale}.  Furthermore, a high purity selection of
dilepton $t\bar{t}$ events can be used to constrain the relative
b-tagging efficiency (Fig.~\ref{fig:top}(right)) to $6 \ (4) \%$ with $1 \
(10) \rm\ fb^{-1}$ of data, as demonstrated in~\cite{topbtageff}.

The electroweak production of single top quarks has been
studied in~\cite{singletop1,singletop2}. Single top production is a
process is sensitive to new physics (e.g. heavy $W'$ bosons, FCNC or charged
Higgs bosons), but also provides a direct handle on the
$|V_{tb}|$ CKM matrix element.  In the t-channel, which has the biggest
cross section, 2400 events are selected with an optimized selection
(${\rm S/B}\sim 1.3$), which allows the cross section to be determined
with an accuracy of $\Delta\sigma/\sigma \sim 2.7 {\rm\ (stat.)} \pm {\rm\ 8.1
(syst.)} \ \%$ for $10 \rm\ fb^{-1}$ of data. The s- and tW-channels
have been investigated as well. There, the estimated uncertainties are
larger.

%%%%%%%%%%%%%%%%%%%%%%%%%%%%%%%%%%%%%%%%%%%%%%%%%%%%%%%%%%%%%%%%%%%%%%%%%%%%%%%

\section{Conclusions}

Due to the large cross sections, the CMS experiment will be able to
make important measurements of $W^\pm,Z^0$ boson and top quark production
already with the first LHC data. These measurements not only constrain
standard model parameters and determine backgrounds to many new
physics signals, but are also very useful as detector commissioning tools and
calibration candles.

% ****************************************************************************
% BIBLIOGRAPHY AREA
% ****************************************************************************

\begin{footnotesize}

\end{footnotesize}

% ****************************************************************************
% END OF BIBLIOGRAPHY AREA
% ****************************************************************************

\end{document}